

\input phyzzx
\indent \hfill{CCNY-HEP-92/1}\break
\indent \hfill January, 1992 \break

\vskip 0.1 in
{\centerline{\bf ONE-DIMENSIONAL FERMIONS AS TWO-DIMENSIONAL DROPLETS}}
{\centerline{\bf VIA CHERN-SIMONS THEORY }} \vskip 0.3 in
\centerline {{\bf Satoshi Iso}\foot{e-mail address:
iso@tkyvax.phys.s.u-tokyo.ac.jp}}
\vskip 0.1 in
\centerline {Uji Research Center, Yukawa Institute for Theoretical Physics}
\centerline{Kyoto University, Uji 611, Japan}
\vskip 0.2 in
\centerline {{\bf Dimitra Karabali}\foot{e-mail address:
karabal@sci.ccny.cuny.edu}{\bf and B. Sakita}\foot{e-mail address:
sakita@sci.ccny.cuny.edu}} \vskip 0.1 in
\centerline{Department of Physics, City College of the City University of New
York}
\centerline{New York, NY 10031}
\vskip 0.3 in
\centerline{ABSTRACT}
\vskip 0.1 in
Based on the observation that a particle motion in one dimension maps to a
two-dimensional motion of a charged particle in a uniform magnetic field,
constrained in the
lowest Landau level,
we formulate a system of one-dimensional nonrelativistic fermions
by using a Chern-Simons field theory in 2+1 dimensions. Using a hydrodynamical
formulation
we obtain a two-dimensional droplet picture of one-dimensional fermions. The
dynamics
involved is that of the boundary between a uniform density of particles and
vortices. We use the sharp boundary approximation. In order to test our
approach we apply
it to a system of fermions in a harmonic oscillator
potential. In the case of well separated boundaries we derive the
one-dimensional collective
field Hamiltonian. Symmetries of the theory are also discussed as properties of
curves in two dimensions.  \pageno=0
\vfill\eject

\Ref\BIPZ{E. Br\'ezin, C. Itzykson, G. Parisi, and J.B. Zuber, {\it Comm. Math.
Phys.} {\bf 59} (1978) 35.}
\Ref\DS{E. Br$\acute{\rm e}$zin, V.A. Kazakov and A.B. Zamolodchikov, {\it
Nucl.Phys.} {\bf B338} (1990) 673; G. Parisi, {\it Phys.Lett.} {\bf B238}
(1990)
209; D. Gross and N. Miljkovi$\acute{\rm c}$, {\it Phys.Lett.} {\bf B238}
(1990)
217; P. Ginsparg and J. Zinn-Zustin, {\it Phys.Lett.} {\bf B240} (1990) 333; J.
Polchinski, {\it Nucl.Phys.} { \bf B346} (1990) 253. }
\Ref\JS{A. Jevicki and B. Sakita, {\it Nucl. Phys.} {\bf B165} (1980) 511.}
\Ref\DJ{S. R. Das and A. Jevicki, {\it Mod. Phys. Lett.} {\bf A5} (1990) 1639.}
\Ref\GK{D. Gross and I. Klebanov, {\it Nucl. Phys.} {\bf B352} (1990) 671.}
\Ref\KS{D. Karabali and B. Sakita, {\it Int. Jour. Mod. Phys.} {\bf A6} (1991)
5079.}
\REF\POL{J. Polchinski, {\it Nucl. Phys.} {\bf B362} (1991) 125.}
\Ref\DDMW{S.R. Das, A. Dhar, G. Mandal and S.R. Wadia, ETH, IAS and
Tata preprint, ETH-TH-91/30, IASSNS-HEP-91/52 and TIFR-TH-91/44, to appear
in Int. J. Mod. Phys. A.; IAS and
Tata preprint, IASSNS-HEP-91/72 and TIFR-TH-91/51, to appear in Mod. Phys.
Lett A.; IAS and Tata preprint, IASSNS-HEP-91/79 and TIFR-TH-91/57.}
\Ref\WINF{J. Avan and A. Jevicki, Brown preprints BROWN-HET-801 and
BROWN-HET-824; M. Awada and
S.J. Sin, Florida preprint UFITP-HEP-91-03; A. Gerasimov, A. Marshakov, A.
Mironov, A. Morozov and
A. Orlov, Lebedev Inst. preprints; D. Minic, J. Polchinsky and Z. Yang, Univ.
of Texas preprint
UTTG-16-91; G. Moore and N. Seiberg, Rutgers and Yale preprint RU-91-29,
YCTP-P19-91; S.R. Das, A. Dhar, G. Mandal and S.R. Wadia, ETH, IAS and
Tata preprint, ETH-TH-91/30, IASSNS-HEP-91/52 and TIFR-TH-91/44, to appear
in Int. J. Mod. Phys. A.; I. Klebanov and
A.M. Polyakov, Princeton University preprint PUPT-1281; E. Witten
IASSNS-HEP-91/51.}
\Ref\GMP{M. Girvin, A. H. MacDonald and P. M.
Platzman, {\it Phys. Rev.} {\bf B33} (1986) 2481.}
\Ref\GM{M. Girvin and A. H. MacDonald, {\it Phys. Rev. Lett.} {\bf 58}
(1987) 1252.}
\Ref\H{C. Hagen, {\it Phys. Rev.} {\bf D31} (1985) 848; R. Jackiw and S. Y. Pi,
{\it Phys.
Rev.} {\bf D42} (1990) 3500.}
\Ref\SSU{B. Sakita, D.-N. Sheng and Z.-B. Su, {\it Phys. Rev.} {\bf B44} (1991)
11510.}
\Ref\TAU{C.H. Taubes, {\it Commun. Math. Phys.} {\bf 72} (1980) 277.}
\Ref\STO{M. Stone, Univ. of Illinois at Urbana preprint IL-TH-91-14.}
\Ref\MO{V. Kazakov, preprint LPTENS 90/30; G. Moore, preprint YCTP-P8-91,
RU-01-12; I.
Klebanov, preprint PUPT-1271.}
\Ref\GP{R.E. Goldstein and D.M. Petrich, {\it Phys. Rev. Lett.} {\bf 67} (3203)
1991.}
\noindent
{\bf I. Introduction}

Nonrelativistic fermions in one space dimension have recently attracted a lot
of attention.
It has been shown [\BIPZ] that they describe the singlet sector of the
one-dimensional hermitian
matrix model, which in the double scaling limit describes the $c=1$ string
model [\DS].
The collective field theory method [\JS] is a powerful
approach to the large $N$ matrix model,
and it has been successfully applied to the string problem [\DJ].
It provides a bosonic field theoretic description of the model and
it can be viewed as a bosonization of apparent
relativistic fermions near the Fermi surface [\GK] or simply as
a bosonization of nonrelativistic fermions [\KS]. In spite of its success, the
method used, namely semiclassical perturbation, is questionable, since
strictly speaking it is not applicable near the turning points.

A rather different approach is the phase space description of Polchinski [\POL]
and the recent
developments [\DDMW] related to it.
Another result, suggestive of the underlying two-dimensional nature of the
problem, is the
emergence of the algebra of area preserving
diffeomorphisms in this context [\WINF] which also appears in the
study of fractional
quantum Hall effect [\GMP]. There should be a close mathematical relationship
between these physically different problems, namely two-dimensional quantum
gravity coupled
to a scalar field and fractional quantum Hall effect.
We like to pursue
this point further in this paper.

Our approach is based on the observation
that a system of fermions in one-space
dimension is equivalent to a system of fermions constrained
in the lowest Landau level. The two-dimensional configuration space of the
latter maps to the phase space of the former.
This is described in section II.

Bosonization in two space dimensions is achieved by
a singular gauge transformation [\GM]
or equivalently by introducing a Chern-Simons gauge field [\H].
Thus we use a model of a
Schr\"odinger Bose field coupled to a Chern-Simons gauge field.
The lowest Landau level condition is imposed by
considering (nonrelativistic)
zero mass charged fermions in a uniform magnetic field.
Since in the Schr\"odinger
theory the mass appears in the denominator of the kinetic energy term,
we first linearize it by using an auxiliary field and then set
the mass to zero. In
effect we obtain a set of second class
constraints describing the lowest Landau level condition.
At this stage one may use the machinery of the Dirac theory of constraints, but
we choose a simpler intuitive approach.
We first solve the constraint equation classically in order to get
fewer unconstrained collective variables, and then develop
a canonical formalism using them.
We extensively use the technique developed in
the collective field approach
to fractional quantum Hall effect [\SSU], which is after all a
hydrodynamical description of fermions.
These issues are discussed in section III.

In section IV we apply our method to a system of free fermions in a harmonic
oscillator potential;
this simple example illustrates the technique and also leads to a
more concrete physical picture.
In this analysis it becomes obvious that the relevant collective variables
are those of a boundary between a uniform density of particles and vortices.
The system becomes
equivalent to a droplet in two dimensions, and the excitations of the system
are due to surface
oscillations.

In section V we extend the theory to the case where the particle density field
occupies a region in
the shape of an infinite strip with the upper and lower boundaries well
separated . We obtain the
collective field Hamiltonian of [\JS] and [\DJ] in a similar manner as in
[\POL]. This is a good
representation of the system provided the potential prevents the two boundaries
from touching. The
separation of boundaries is achieved in the harmonic oscillator potential, but
not in the inverse
harmonic oscillator potential with small chemical potential.

The final section is devoted to
discussions in which we also include
symmetries and a space-time description of the theory.

\noindent
{\bf II. The Relation Between Particles in One Space Dimension and Particles in
Two Space Dimensions in Strong Magnetic Field}

Let us consider a particle in a potential in one space dimension. The
Hamiltonian is given by
$$
{H = {1\over {2m}} p^2 + v(x) .}\eqno (2.1)
$$
For the specific case of harmonic oscillator, which we shall investigate
extensively below, the potential is given by
$$
{v(x)={m\over 2 } \omega^2 x^2 .}\eqno (2.2)
$$
In quantum theory $p$ is a differential operator given by $-i{\partial\over
{\partial x}}$.

Next let us consider a massless charged particle in a uniform magnetic field
in two space dimensions. The Lagrangian is given by
$$
{L={\bf A }\cdot {\dot{\bf x}} - A_0 (x, y) .}\eqno (2.3)
$$
In Landau gauge the vector potential ${\bf A}$ is given by
$$
{{\bf A}\equiv ( yB , 0) , ~~~~~~~~{\rm with}~~ B>0 }\eqno (2.4)
$$
so that the first term of $L$ is
$$
{{\bf A}\cdot  \dot{\bf x}=   B
y{\dot x} .} \eqno (2.5)
$$
Since the canonical momentum of $ x $ is
$$
{p_x = By ,}\eqno (2.6)
$$
the Hamiltonian of the system can be written as
$$
{H= A_0 (x, {1\over B}p_x ) .}\eqno (2.7)
$$
Thus the one dimensional particle system is equivalent to a two dimensional
massless charged particle system under uniform magnetic field provided
$$
{A_0 (x,y)={B^2\over{2m}}y^2 + v(x) .}\eqno (2.8)
$$

In the following section we shall exploit this observation for systems of
many fermions in one space dimension and develop a theory of Fermi liquid by
using the
field theory of Chern-Simons gauge field coupled to a nonrelativistic
Schr\"odinger Bose
field in two space dimensions.

\noindent
{\bf III. Chern-Simons Field Theory for Massless Fermions}

Let us first consider a system of massive (mass $m_0$) fermions in
electromagnetic
field in two space dimensions. The Hamiltonian  is given by
$$\eqalignno{
H&={1\over{2m_0}}\sum_{a=1} ^N ( {\bf \Pi }_a )^2 + \sum_a A_0
({\bf x}_a )\cr &={1\over{2m_0}}\sum_{a=1} ^N ( \Pi^x _a +i\Pi^y _a )
( \Pi^x _a -i\Pi^y _a )+ \sum_a A_0 ({\bf x}_a ) +{B\over{2m_0}}N , &(3.1)\cr}
$$
where
$$
\Pi^i =  p_i -A^i ({\bf x}) =-i {\partial \over \partial x^{i}}-A^i ({\bf x}),
\ \ \ \
\  \ \ \ \  i= x, y \eqno (3.2)
$$
We use the metric $( +,-,-)$.
We bosonize the above quantum mechanical system by using a singular gauge
transformation [12].
As a result we obtain an effective bosonic Hamiltonian. Its form is the
same as in (3.1) except that $\Pi^i$ is now given by $$
\Pi^i _a = p_{ia} -A^i ({\bf x_a}) - a^i ({\bf x_a}) ,\eqno (3.3)
$$
where $a^i ({\bf x_a})$ is
$$
a^i ({\bf x_a})= -\epsilon^{ij}\sum_{b\ne a}
{{(x_a - x_b )^j}\over{|{\bf x}_a -{\bf x}_b
|^2}} .\eqno (3.4)
$$
The effective bosonic system can be described in a second quantized language in
terms of a
Bose Schr\"odinger wave field $\psi$ [\KS]. We further introduce a Chern-Simons
gauge field as an
auxiliary field to ensure (3.4). We obtain the following Lagrangian density for
the original
fermionic system
$$ {{\cal L} = \bar\psi \Pi_0\psi -{1\over{2m_0}}\bar\psi (\Pi^x +i\Pi^y )
(\Pi^x
-i\Pi^y )\psi +{\cal L}_{CS},}\eqno (3.5) $$
where
$$
{\Pi_{0} = i\partial _0 - A_{0} - a_{0},
\ \ \ \ \ \ \ \ {\cal L}_{CS} =
-{1\over{4\pi}}\epsilon^{\mu\nu\rho}a_{\mu}\partial_{\nu}a_{\rho} .}\eqno (3.6)
$$

Next we linearize the kinetic energy term by using auxiliary fields $\lambda
,\bar\lambda$:
$$
{{\cal L} =\bar\psi \Pi_0\psi -\bar\psi (\Pi^x +i\Pi^y )\lambda -\bar\lambda
(\Pi^x -i\Pi^y )\psi +{\cal L}_{CS} +2m_0|\lambda |^2}\eqno (3.7)
$$
and then set
$m_0 =0$.
The auxiliary field $\lambda$ plays the role of a Lagrange multiplier.

Next we change variables from $\psi , \bar \psi$  to $\rho , \theta$ through $$
{\psi =\sqrt{\rho} e ^{i\theta}} . \eqno (3.8)
$$
We obtain
$$
{{\cal L} = b_{\mu}(\partial^{\mu}\theta + A^{\mu} +a^{\mu}
-{1\over 2}\epsilon^{\mu 0\nu}\partial_{\nu} \ln\rho ) + {\cal L}_{CS} ,
}\eqno (3.9)$$
where
$$
{b^0 = -\rho ,\ \ \ \ \ \ \ \  {1\over 2} ( b_x + i b_y )
= \bar\lambda\psi} . \eqno (3.10)
$$
The phase $\theta$ can have a regular
part and a singular part :
$$
{\theta = \theta_{\rm reg} +\theta_s} .\eqno (3.11)
$$
The singular part is due to a vortex configuration:
$$
{\partial^{\mu}\theta_s = v^{\mu}, \ \ \ \ \  \nabla\times{\bf v} =
2\pi\rho_V ,  \ \ \ \ \ \nabla\cdot{\bf v}=0} \eqno (3.12)
$$
where $\rho_V$ is the density of vortices.
If we set $$
{b^{i} = \epsilon^{ij}\partial_{j} h + \partial^{i} g} \eqno (3.13)
$$
and insert this into (3.9) we see that $a_{0},~h$ and $g$ act as Lagrange
multipliers. We obtain
$$
{{\cal L}=-\rho(\dot{\theta} _{reg} + v_{0} + A_{0}) + {1 \over 4 \pi}
\epsilon^{ij} a_{i}
\dot{a}_{j}} \eqno(3.14)
$$
along with the conditions imposed by the Lagrange multipliers
$$\eqalignno
{& \nabla\times{\bf a}=2 \pi \rho \cr
& \nabla\times{\bf v}  +\nabla\times{\bf A}
+\nabla\times{\bf a}- {1\over 2}\nabla^2\ln\rho
 =0\cr
& \nabla ^{2} \theta _{reg} + \nabla\cdot {\bf a}=0 &(3.15) \cr}
$$
Using the fact that $a_{i}$ can be written as
$$
a_{i}={1 \over {\nabla^2}}\left( 2 \pi \epsilon_{ij} \partial_{j} \rho -
\partial _{i} \nabla \cdot
{\bf a}\right) \eqno(3.16) $$ we notice that the first and last term in
(3.14) cancel each other because of (3.15).
Therefore,
the dynamics of the system is simply described by
$$
{{\cal L} = -(v_0 + A_0 )\rho}\eqno (3.17)
$$
with the subsidiary condition (notice that $\nabla \times {\bf A}=-B,~~B>0$)
$$
{B + {1\over 2}\nabla^2\ln\rho -2\pi\rho_V
-2\pi\rho =0}. \eqno (3.18)
$$
This subsidiary condition is the same as the constraint equation for
electrons being in the lowest Landau level in the fractional quantum
Hall system [\SSU].

Notice that
$$
- \int d^2 x v_0\rho =  \int\int \rho (x) 2\pi G(x-x^{\prime}) \dot{\rho}_V
(x^{\prime}) d^2 x d^2 x' \eqno(3.19)
$$
where $G$ is the Green's function
satisfying  $$
{\epsilon^{ij}\partial_i\partial_j G(x-x') = \delta^2 (x-x'),
\ \ \ \ \ \ \ \ \nabla^2 G=0}.\eqno (3.20)
$$
An explicit form of $G$ is given by
$$
{G(x)={\theta\over {2\pi}} = {{Im\ln z }\over{2\pi}}}.\eqno (3.21)
$$

We first solve the subsidiary condition (3.18)
and then develop a canonical formalism.

The uniform field configuration
$$
\rho={B\over {2\pi}}, \ \ \ \ \ \ \ \ \rho_V =0\eqno (3.22)
$$
is the
simplest field configuration for which the subsidiary condition (3.18) is
satisfied. This corresponds to a uniform distribution in the
phase space of one dimensional fermions. It is obviously
unphysical since $A_0$
given by (2.8) makes the
second term in the Lagrangian (3.17), i. e. the energy, infinite.
\footnote*{This state is
perfectly physical for the fractional quantum Hall system. It corresponds to
the case where all
the lowest Landau levels are completely filled.} One should select solutions of
the subsidiary
condition in such a way that the energy $\int d^2 x A_0 \rho $ stays in the
neighborhood of
minimum. Thus  the canonical procedure suggested earlier depends on the form of
$A_0$. We shall
discuss it in detail for the case of a harmonic oscillator potential (2.2) in
the
following section.

\noindent
{\bf IV. Fermions in Harmonic Oscillator Potential}

Here we study a system of $N$ fermions in a harmonic oscillator potential.
The single particle Hamiltonian is given by (2.1) and (2.2). The system
consists of $N$ fermions occupying the equally spaced energy levels. The ground
state and the excited states are easily obtained by the standard independent
particle picture. It is known that in the large $N$ limit the low frequency
excitations are classified by
using a set of Bose oscillators with frequency $\omega_n = n\omega ,
\ \ (n=1,2,\cdots ,)$.
We shall
describe this simple known problem by using the method described in the
previous section.

Let us choose
$$
B=m\omega \eqno (4.1)
$$
in (2.8) and combine it with (2.2). Then we
obtain
$$\eqalignno
{A_0 = &{1\over 2} m\omega^2 (x^2 + y^2 ) -\mu \cr
= &
{1\over 2} m\omega^2 ((x^2 + y^2 ) - r_0 ^2 ), & (4.2) \cr}
$$
where $\mu$ is a chemical potential and we parametrize it as in the second
line. It is quite obvious that the lowest energy configuration is the one where
$\rho$ takes the
maximum value inside a circle of radius $r_0$ and the minimum value outside.

We notice that in general the constraint equation (3.18) is the equation for
vortices. If one appropriately scales the coordinates and then sets $4\pi\rho =
e^{2f_1}$, one obtains
$$
-\Delta f_1 +{1\over 2}(e^{2f_1} -1) = - 2\pi \rho_V .
\eqno (4.3)
$$
Along with the requirement of singlevaluedness and finiteness of $\psi$, one
finds [\TAU] that (4.3) admits vortex solutions if
$\rho_V$ is given by $$
\rho_V = \sum_k n_k \delta^{(2)} (x - x_k ) , \eqno (4.4)
$$
where ${\bf x}_k$'s are vortex positions and the vorticities, $n_k$'s, are {\it
positive}
integers. The particle density $\rho$ is {\it zero} at the position of vortex.

Coming back to the harmonic oscillator potential problem,
in order that the energy stays in the neighborhood of the minimum, we must
choose configurations with almost no vortices
inside the circle and infinitely many, densely populated outside. Below we
shall use, for
solutions of the subsidiary condition (3.16), the sharp boundary approximation
where $\rho=B/2 \pi$
inside and zero outside the boundary and respectively $\rho _{V} =B/2 \pi$
outside and zero inside
the boundary. We parametrize the boundary by $\delta r(\theta )$ and treat it
as a dynamical
variable. $$ \eqalignno{\rho (x) \ & = {B\over {2\pi}}\theta (-r +r_0 + \delta
r (\theta ) )\cr
& \approx {B\over{2\pi}} ( \theta ( - r + r_0 ) +\delta r(\theta )\delta
(r - r_0 ) - {{(\delta r )^2 }\over 2} \delta ' ( r - r_0 ) + \cdots ) &
(4.5)\cr {{\rho}}_V (x) & =
{B\over {2\pi}}\theta (r - r_0
- \delta r (\theta ) )\cr
 & \approx {B\over{2\pi}} ( \theta (  r - r_0 ) -\delta r(\theta
)\delta (r - r_0 ) + {{(\delta r )^2 }\over 2} \delta ' ( r - r_0 ) + \cdots )
& (4.6)  \cr}
$$
Since the particle number $N$ is proportional to $r_0 ^2$, this expansion is
a large $N$
expansion.

For the first term in the Lagrangian (3.17) we obtain
$$
{- \int \rho v_0 =-r_0 ^2 \left({B\over {2\pi}}\right)^2 \int \int \delta r
(\theta ) Im \ln ( e^{i\theta} - e^{i\theta '})\delta {\dot r} (\theta ' )
d\theta d \theta ' }, \eqno (4.7)
$$
keeping up to the quadratic term in $\delta r$ (higher order terms are with
higher power of $r_0^{-1}$).

Expanding $\delta r$ as
$$
{\delta r (\theta ) = {{\sqrt 2 }\over { r_0 B}}\sum_{n>0}
\sqrt{n} ( q_n \sin n\theta + p_n \cos n\theta )}\eqno (4.8)
$$
and noticing that
$$
\eqalignno{Im \ln (e^{i\theta}-e^{i\theta'}) & =\theta-Im \sum_{n>0}
{e^{in(\theta'-\theta)} \over n}\cr
  &=\theta-\sum_{n>0}{1\over n}\left(\sin n\theta' \cos n\theta -
\cos n\theta' \sin n\theta\right)&
 (4.9) \cr}$$
we obtain that
$$
{- \int \rho v_0 =  \sum_{n>0} p_n {\dot {q}}_n}\eqno (4.10)
$$
up to a total derivative,
and
$$
{\int A_0\rho =\sum_{n>0} n\omega {1\over 2}(p_n ^2 + q_n ^2 )}\eqno (4.11)
$$
up to a constant.
Thus the system is equivalent to an ensemble of harmonic oscillators.
We obtain the collective excitation spectrum
$$
\omega_n = n\omega , \eqno (4.12)
$$
which we anticipated.

\noindent
{\bf V. Reduction To A Field Theory In One Space
Dimension}

Guided by the above analysis, we formulate, in this section, the theory for a
general case in which
the particle density is non-zero within two well separated boundaries.
We take the following expression as an approximate
solution of the subsidiary condition
(3.18):
$$
\rho (x, y, t)={B \over 2\pi} \theta(y_+ (x,t) -y) \theta(y-y_- (x,t)) \eqno
(5.1) $$
We assume that the time variation of $y_{\pm}$ is much smaller than the
difference $y_+ - y_-$ and set
$$
y_{\pm} (x,t) = y_{\pm} ^0 (x) +\delta y_{\pm} (x,t),\ \ \ \ \ \ \ \ \
|\delta y_{\pm} (x,t)| \ll y_+ ^0 (x) - y_- ^0 (x) \eqno (5.2)
$$
We then expand $\rho$ and $\rho_V$ as
$$
\eqalign{\rho &\approx {B \over 2\pi}\left(\theta(y_{+}^0 -y) \theta(y-y_-^{0}
)
+\delta y_{+}\delta(y-y_{+}^0
) -\delta y_{-} \delta(y-y_{-}^0 ) +\cdots \right) \cr
{{\rho}_V} &\approx {B \over 2\pi}\left(\theta(y- y_{+}^0 ) + \theta(y_-^{0}
-y ) -\delta y_{+}\delta(y-y_{+}^0
) +\delta y_{-} \delta(y-y_{-}^0 ) +\cdots\right)\cr}\eqno (5.3)
$$
and insert the above expressions into the first term in the Lagrangian. We
obtain
$$
\eqalignno{ -\int\rho v_0 =-\left( {B \over 2\pi}
\right)^2 \int  dx dx' \bigl( &
\delta y_{+}(x)\delta \dot y_{+}(x') Im \ln(x+iy_+ ^0 (x) - x' -iy_+ ^0 (x') )
\cr &+\delta y_{-}(x)\delta \dot y_{-}(x') Im \ln(x+iy_- ^0 (x) - x' -iy_-
^0 (x') )\bigr)  \cr
& & (5.4)\cr}
$$
The cross term has been omitted in the above expression since it can be shown
to have
a form of total derivative in time.

The Green's function is in general written as
$$
Im \ln(x+iy_\pm ^0 (x) - x' -iy_\pm ^0 (x')) = \pm \pi \theta(x-x')
+ f_{\pm} (x, x'), \eqno
(5.5) $$
where $ f_{\pm} (x, x')$ is symmetric under the interchange of $x $ and $x'$
and therefore
it does not contribute to (5.4). In deriving (5.5) we took into account the
fact
that the prime coordinate corresponds to the vortices, therefore it lies above
the
upper boundary and below the lower boundary. Thus (5.4) is now given by
$$
-\int\rho v_0 =-\left( {B \over 2\pi}
\right)^2 \pi \int  dx dx' \bigl( \delta y_{+}(x)\delta \dot y_{+}(x') -
\delta y_{-}(x)\delta \dot y_{-}(x')\bigr)\theta (x-x') .\eqno (5.6)
$$

The normal mode expansion is given by
$$
\delta y_\pm (x) =\int^{\infty}_{0} dk \left( \delta y_\pm ^R (k)\,\cos  kx  +
\delta y_\pm ^I (k)\,\sin
 kx  \right) . \eqno (5.7)
$$
Using that $\theta (x) = {1 \over {2 \pi i}} \int du {{e^{iux}} \over {u-ie}}$,
we obtain
$$
-\int\rho v_0 = -{B^2  \over 2}  \int _{0}^{\infty} dk {{\delta y_+ ^I (k)
\delta\dot y_+
^R (k)- \delta y_- ^I (k) \delta\dot y_- ^R (k)} \over k} . \eqno (5.8)
$$
where we discarded again a total time derivative term.
Thus the quantum commutation relations are given by
$$
[\delta y_\pm ^R (k), \delta y_\pm ^I (k') ] =\mp i {{2 k} \over{B^2}} \delta
(k-k') , \eqno
(5.9) $$
accordingly
by
$$
[y_\pm (x) , y_\pm (x') ] =\mp i{2\pi\over{B^2}}\delta ' (x-x') .\eqno (5.10)
$$

The Hamiltonian is given by
$$
\eqalignno{
    \int \rho A_0 & =\int \rho({B^2 y^2 \over 2}+v(x)-\mu) d^2 r \cr
    &={B \over 2\pi} \int dx \left( {B^2 \over 6}\left( y_{+} ^3 (x) - y_{-}
^3 (x)\right)  +(v(x)-\mu)
     \left( y_{+} (x)- y_{-}(x)\right) \right) .
         & (5.11) \cr }
$$
where we set $m=1$.
Let us define
$$
y_\pm (x) = {1\over B} (\pm \pi \phi (x) + \partial \Pi (x) )\eqno (5.12)
$$
where $\phi$ and $\Pi$ satisfy the standard canonical commutation relation. In
terms of these new
variables the Hamiltonian is given by
$$
\int \rho A_0  =\int dx \left( {1\over 2}\phi (\partial \Pi )^2
+ {{\pi^2}\over 6}\phi^3 +(v(x)-\mu)\phi\right)\eqno (5.13)
$$
which is precisely the collective field Hamiltonian [\JS,\DJ].

\noindent
{\bf VI. Discussions}

We have formulated one-dimensional nonrelativistic fermions by using
Chern-Simons field theory. After converting the field variables
into the hydrodynamical collective variables and eliminating the constraint, we
reached a droplet picture of one-dimensional fermions. It is a system of
incompressible liquid in two dimensions. The dynamical degrees of freedom are
the fluctuations of
the boundary between a uniform density of particles and vortices. These are
similar to the edge
excitations described in [\STO].

In section V we obtained the collective field
Hamiltonian for the case of a
boundary curve that does not fluctuate widely, so it can be described as a
single valued function of $x$, and further more it does not touch itself.
Obviously this
should be modified for general cases.  In the case that the curve touches
itself one has
to take into account possible rearrangement of the boundaries.

These issues are most relevant in the case of an inverted harmonic oscillator
potential with
small chemical potential. This corresponds to the $c=1$ string model with
strong coupling
[\MO]. It
is clear from Fig.1 that rearrangement of the boundary is necessary when one
considers
boundary oscillations near the turning points.

We should mention that the one-dimensional collective field Hamiltonian derived
here differs from the one presented in [\KS], where we directly bosonized the
one-dimensional fermions, by a subleading in $N$ order term, which contains
derivatives of the particle density. We believe that the sharp boundary
approximation used is responsible for the absence of this term. Further in the
case of the inverted harmonic oscillator potential with small chemical
potential,
appropriate treatment of the surface rearrangement in the region of the
turning points will produce the effect of the extra derivative term in the
collective field Hamiltonian, since they both express a tunneling effect.

In general in order to accommodate a possible rearrangement of the boundaries
we
follow the surface motion in space-time and consider a connected region
$\Omega$
in Minkowski space, in which the liquid flows. We choose $\Omega$ as shown in
Fig.2 to include fission and fusion of droplets. The action is given by
$$
S=-\int_{\ \ \Omega}(v_0 + A_0 ) \rho.\eqno (6.1) $$
This is essentially (3.17), but
here $\rho ({\bf x},t)$ is $1$ (we set $B/2\pi =1$) inside $\Omega$ and zero
outside and ${\rho} _V$ is the other way around. The boundary of $\Omega$ is
specified by  $$ x^\mu =X^\mu (\sigma ,\tau).\eqno (6.2) $$ Thus $S$ is a
functional of $X^\mu (\sigma ,\tau)$.  Based on this action, in principle, one
can discuss the quantum mechanical motion of surfaces. We hope we can come back
to this problem in future.

We shall now consider briefly the symmetry of the system.
It is obvious that reparametrization of the surface is a symmetry.
The issue is whether this symmetry is relevant to the dynamics of surfaces.
Since the functional dependence of $X^\mu (\sigma ,\tau )$ is implicit in the
action (6.1),  we look at the
classical equation of
motion. For this purpose we consider a droplet with a simple surface
and
parametrize it by setting $\tau =t$, where $t$ is time.
The boundary of the droplet at $t$ is given by a two dimensional closed
curve ${\bf r}(s , t)$ . In this discussion we use ${\bf r}(s)$
instead of ${\bf X}(\sigma )$ in order to conform to the notation of [\GP]. The
density $\rho ({\bf x}, t) $ is expressed as
$$
\rho ({\bf x},t) =\int_{\ \ \Omega (t)} d{\bf x}'\delta ({\bf x}-{\bf x}'),
\eqno (6.3)
$$
and its variation with respect to a boundary change is given by
$$
\delta \rho ({\bf x}, t) =\oint ds \, \delta{\bf r}\times{\bf r}_s\,
\delta ({\bf x}-{\bf r}(s , t)),\eqno (6.4)
$$
and a similar expression for ${\rho}_V$. The variation of action (6.1) is then
given by
$$ \eqalignno
{\delta S = &\int dt \oint ds \oint ds'\, {\bf r}_s (s,t)\times\delta
{\bf r}(s,t) \cr
&\ \ \ \ \ \ \ \ \left( Im\ln (z_+ (s) -z_- (s')) -Im\ln (z_+ (s') -z_- (s)
\right) {\bf r}_s (s',t)\times {\bf r}_t (s',t)\cr
&\ \ \ \ \ \ \ \ \ \ \ \
\ \ \  +\int dt \oint ds \, {\bf r}_s (s,t)\times\delta {\bf r}(s,t) A_0
({\bf r}(s,t)),&(6.5)\cr} $$
where $z_+ (s),\, z_-(s)$ are points infinitesimally close to the boundary
$z(s)$, inside and
outside respectively. Since
$$
Im\ln (z_+ (s) -z_- (s')) -Im\ln (z_+ (s') -z_- (s)) =\pi\epsilon (s-s')
{}~~~~{\rm mod}(2
\pi) \eqno(6.6) $$
where
$$\epsilon (s-s')=\cases{+1 & for  $s-s' >0$ \cr  -1 & for
$s-s' <0$ \cr} $$
we obtain the following equation of motion:
$$
2\pi {\bf r}_t \times{\bf r}_s ={\partial\over{\partial s}}A_0 ({\bf r}(s,t)).
\eqno (6.7)
$$
This equation is invariant under {\it time dependent} reparametrization of $s$.
It is amusing to notice that the enclosed area is a constant of motion as
a consequence of (6.7), since
$$
{\partial\over{\partial t}}\int d {\bf x} \rho ({\bf x},t) =
\oint ds {\bf r}_t\times{\bf r}_s ={1 \over 2\pi}\oint ds
{\partial\over{\partial s}}A_0 ({\bf r}(s,t)) =0 .\eqno (6.8)
$$
After all it merely states the fermion number conservation.

In a recent paper, Goldstein and Petrich [\GP]
discuss
the relation between the KdV hierarchy and closed curve dynamics of
the form
$$
{\bf r}_t = U{\hat{\bf n}} + W{\hat{\bf t}}, \eqno (6.9)
$$
where {\bf n} and {\bf t} are normal and tangential unit vectors respectively.
By a suitable reparametrization of $s$ one can set ${\bf t}={\bf r}_s $.
Thus,
the dynamics of our curve corresponds to that of
$$
U={1 \over {2 \pi}}{\partial\over{\partial s}}A_0 ({\bf r}),
$$
a total derivative in $s$ , and {\it arbitrary} $W$. Since $W$ is
arbitrary, this includes a much wider class than the KdV.

\noindent
{\bf Acknowledgements}

We acknowledge Spenta Wadia, whose critical comment prompted us to consider
the present work. The first author acknowledges the last for the hospitality
extended to him at City College. This work is supported by the NSF grant
PHY90-20495 and the Professional Staff Congress Board of Higher Education of
the City
University of New York under grant no. 6-62356.

\refout

\vfill\eject
\centerline{FIGURES}

\vskip 0.5 in
FIG.1.  Rearrangement of the boundary.
\vskip 0.3 in
FIG.2.  Fission and fusion of droplets.

\end